\documentclass[journal]{IEEEtran}
\usepackage{cite}

\usepackage{color,soul}
\usepackage{gensymb}
\usepackage{dsfont}

\ifCLASSINFOpdf
  \usepackage[pdftex]{graphicx}
  \graphicspath{{../pdf/}{../jpeg/}}
  \DeclareGraphicsExtensions{.pdf,.jpeg,.png}
\else
  \usepackage[dvips]{graphicx}
  \graphicspath{{../eps/}}
  \DeclareGraphicsExtensions{.eps}
\fi
\usepackage[cmex10]{amsmath}
\usepackage{amssymb}

\usepackage{euscript}

\DeclareMathAlphabet{\mathpzc}{OT1}{pzc}{m}{it}
\usepackage{multirow}
\usepackage{algorithm}
\usepackage[noend]{algpseudocode}
\usepackage{diagbox}
\usepackage{physics}
\usepackage{tikz}

\ifCLASSOPTIONcompsoc
  \usepackage[caption=false,font=normalsize,labelfont=sf,textfont=sf]{subfig}
\else
  \usepackage[caption=false,font=footnotesize]{subfig}
\fi

\usepackage{url}

\begin{document}
%
\title{Statistical Modeling of Networked Solar Resources for Assessing and Mitigating Risk of Interdependent Inverter Tripping Events in Distribution Grids
}
%
%
%

\author{
         Kaveh Dehghanpour,~\IEEEmembership{Member,~IEEE,}
         Yuxuan Yuan,~\IEEEmembership{Student Member,~IEEE,}
         Fankun Bu,~\IEEEmembership{Student Member,~IEEE}
         Zhaoyu Wang,~\IEEEmembership{Member,~IEEE}
         
\thanks{This work was supported by the Advanced Grid Modeling Program at the U.S. Department of Energy Office of Electricity under Grant DE-OE0000875. (\textit{Corresponding author: Zhaoyu Wang})

K. Dehghanpour, Y. Yuan, F. Bu, and Z. Wang are with the Department of
Electrical and Computer Engineering, Iowa State University, Ames,
IA 50011 USA (e-mail: kavehd@iastate.edu; wzy@iastate.edu).
}
}
%
%

\markboth{Submitted to IEEE for possible publication. Copyright may be transferred without notice}%
{Shell \MakeLowercase{\textit{et al.}}: Bare Demo of IEEEtran.cls for Journals}
%



\maketitle

\begin{abstract}
It is speculated that higher penetration of inverter-based distributed photo-voltaic (PV) power generators can increase the risk of tripping events due to voltage fluctuations. To quantify this risk utilities need to solve the interactive equations of tripping events for networked PVs in real-time. However, these equations are non-differentiable, nonlinear, and exponentially complex, and thus, cannot be used as a tractable basis for solar curtailment prediction and mitigation. Furthermore, load/PV power values might not be available in real-time due to limited grid observability, which further complicates tripping event prediction. To address these challenges, we have employed Chebyshev's inequality to obtain an alternative probabilistic model for quantifying the risk of tripping for networked PVs. The proposed model enables operators to estimate the probability of interdependent inverter tripping events using only PV/load statistics and in a scalable manner. Furthermore, by integrating this probabilistic model into an optimization framework, countermeasures are designed to mitigate massive interdependent tripping events. Since the proposed model is parameterized using only the statistical characteristics of nodal active/reactive powers, it is especially beneficial in practical systems, which have limited real-time observability. Numerical experiments have been performed employing real data and feeder models to verify the performance of the proposed technique.    
\end{abstract}


%
\IEEEpeerreviewmaketitle

\section{Introduction}
Increasing penetration of distributed energy resources (DERs), including inverter-based photo-voltaic (PV) power generators, in distribution grids represents opportunities for enhancing system resilience and customer self-sufficiency, as well as challenges in grid control and operation. One of these challenges is the potential increase in the risk of tripping of inverter-based resources due to undesirable fluctuations in the grid's voltage profile \cite{Tonkoski2012}. This can put a hard limit on the feasible capacity of operational PVs in distribution grids, reduce the economic value of renewable resources for customers, and cause loss of service in stand-alone systems \cite{Gagrica2015,Hasheminamin2015}. The possibility of DER power generation disruption due to voltage-related vulnerabilities in unbalanced distribution grids has been discussed in the literature: in \cite{Ferreira2013,Carvalho2016}, risk of interdependent tripping of PVs, with ON/OFF current interruption mechanism was demonstrated numerically in a distribution grid test case for the first time. It was shown that the unbalanced and resistive nature of networks can further exacerbate this problem by causing positive inter-phase voltage sensitivity terms that act as destabilizing positive feedback loops, leading to voltage deviations after tripping of an individual inverter. The impact of grid voltage sensitivity on DER curtailment was also studied and observed in \cite{Gagrica2015}. Based on these insights, guidelines were provided in \cite{Carvalho2015} to roughly estimate the impact of new DER capacity connections on the maximum voltage deviations in the grid. It was shown in \cite{Popiel} that very large or small number of inverter-based resources in distribution systems can lead to interdependent failure events that contribute to voltage collapse in transmission level. Detailed realistic numerical studies were performed on practical feeder models in \cite{Parchure2017, Cheng2016, Gupta, Ding2017, Xu2018, Jiang2015} that corroborated the considerable impacts of extreme PV integration levels, and inverter control modes on grid voltage fluctuations, which is the critical factor in causing massive solar curtailment scenarios.

Most existing works relied on scenario-based simulations and numerical studies to capture the likelihood of inverter tripping under high renewable penetration. While this has led to useful guidelines and invaluable intuitions, it falls short of providing a generic theoretical foundation for predicting and containing tripping events. Specifically, the dependencies between nodal solar power distributions, nodal voltage profiles, and inverter tripping events have not been explicitly analyzed in the literature thus far. These dependencies are influenced by inverter protection settings and governed by a set of networked power-flow-based equations, which turn out to be non-differentiable and nonlinear. In this regard, several fundamental challenges have not been addressed: (1) \textit{Lack of scalability:} Solving the inverter tripping equations directly in real-time requires a large-scale search process to explore almost all the joint combinations of ``ON/OFF'' configurations for the inverters. The computational complexity is due to the interactive and networked nature of tripping events, meaning that the states of inverters influence each other and are not independent \cite{Jennett2014}. The source of interdependency in chances of inverter tripping is the dependencies in nodal voltages of power grid (i.e., disruption of power injection at one node impacts nodal voltages of other neighboring inverters, which in turn could influence their probability of tripping.) For example, tripping of an inverter (or a cluster of inverters) leads to a change in loading distributions, which can most likely increase/decrease the chance of tripping for other inverters during under/over-voltage scenarios, especially in weak grids. This interdependency prevents the solver from decoupling the tripping equations into separate equations for individual inverters. Thus, the scale of search for finding the correct configuration increases exponentially ($2^N$) with the number of inverters ($N$). Another factor that contributes to computational complexity is the volatility of PV power, which forces the solver to explore, not only various tripping configurations, but also numerous solar scenarios at granular time steps. (2) \textit{Limited tractability for mitigation:} A direct solution strategy for tripping equations cannot be easily integrated into optimization-based decision models, since it has no predictive capability and cannot be use to answer \textit{what if} queries, unless a thorough expensive search is performed over all possible future load/PV scenarios. Also, due to their non-differntiability, integrating the tripping equations into decision models complicates formulation by adding integer variables to the problem. (3) \textit{Limited access to online data:} Practical distribution grids have low online observability, meaning that the values of real-time nodal power injections can be unknown in real-time for a large number of PVs/loads due to communication time delays or limited number of sensors. Thus, we might not have access to sufficient online information to solve the tripping problem directly.

To tackle these challenges, we propose an alternative probabilistic modeling approach to quantify and mitigate the risk of voltage-driven tripping events. Instead of complex scenario-based look-ahead search over numerous possible tripping configurations, our methodology is built upon probabilistic manipulation of power flow equations in radial networks to estimate the probability of inverter tripping using only the available statistical properties of loads/PVs. Interdependent Bernoulli random variables are used to model probabilities of inverter tripping and capture their mutua. These probabilities are voltage-dependent and serve as unknown \textit{micro-states} in the equations of tripping events. Then, Chebyshev's inequality \cite{Ross2006} is applied to determine a stationary lower bound for the values that these micro-states can assume under any probable nodal power injection scenarios. This lower bound provides a conservative estimation of expected PV curtailment, and thus, represents a statistical risk metric for tripping events. Furthermore, due to its simple matrix-form and differentiable structure, the proposed probabilistic model can be conveniently integrated into an optimization framework as a constraint, which enables mitigating unwanted solar curtailment events by designing optimal voltage regulation countermeasures. The proposed methodology is generic and can capture the behavior of arbitrary radial distribution feeders using only load/PV statistics and network topology/parameters. This implies that tripping events can be conservatively predicted using the proposed model and without the need for online access to granular PV/load data or expensive scenario-based search process, which makes our strategy specifically suitable for practical networks. 

Numerical experiments have been performed using real advanced metering infrastructure (AMI) data and feeder models from our utility partners to validate the developed probabilistic framework. The numerical validates the performance of the probabilistic model for both over- and under-voltage scenarios, and show that ignoring the possibility of tripping in voltage regulation can exacerbate voltage deviations.   


\section{Deriving A Conservative Probabilistic Model of PV Tripping Events}\label{sec:dyn}
In this section, we will develop and then parameterize a probabilistic model of networked inverter-based PVs to quantify the possibility of emergent tripping events. To do this, first, we begin with the original model of inverter tripping with ON/OFF voltage-driven current interruption mechanism, and then, we will show that by adopting a probabilistic approach towards the original model and using Chebyshev's inequality, tripping probabilities can be conservatively estimated using the statistical properties of nodal available load/PV power. 
\begin{figure}
\centering
\includegraphics[width=0.9\columnwidth]{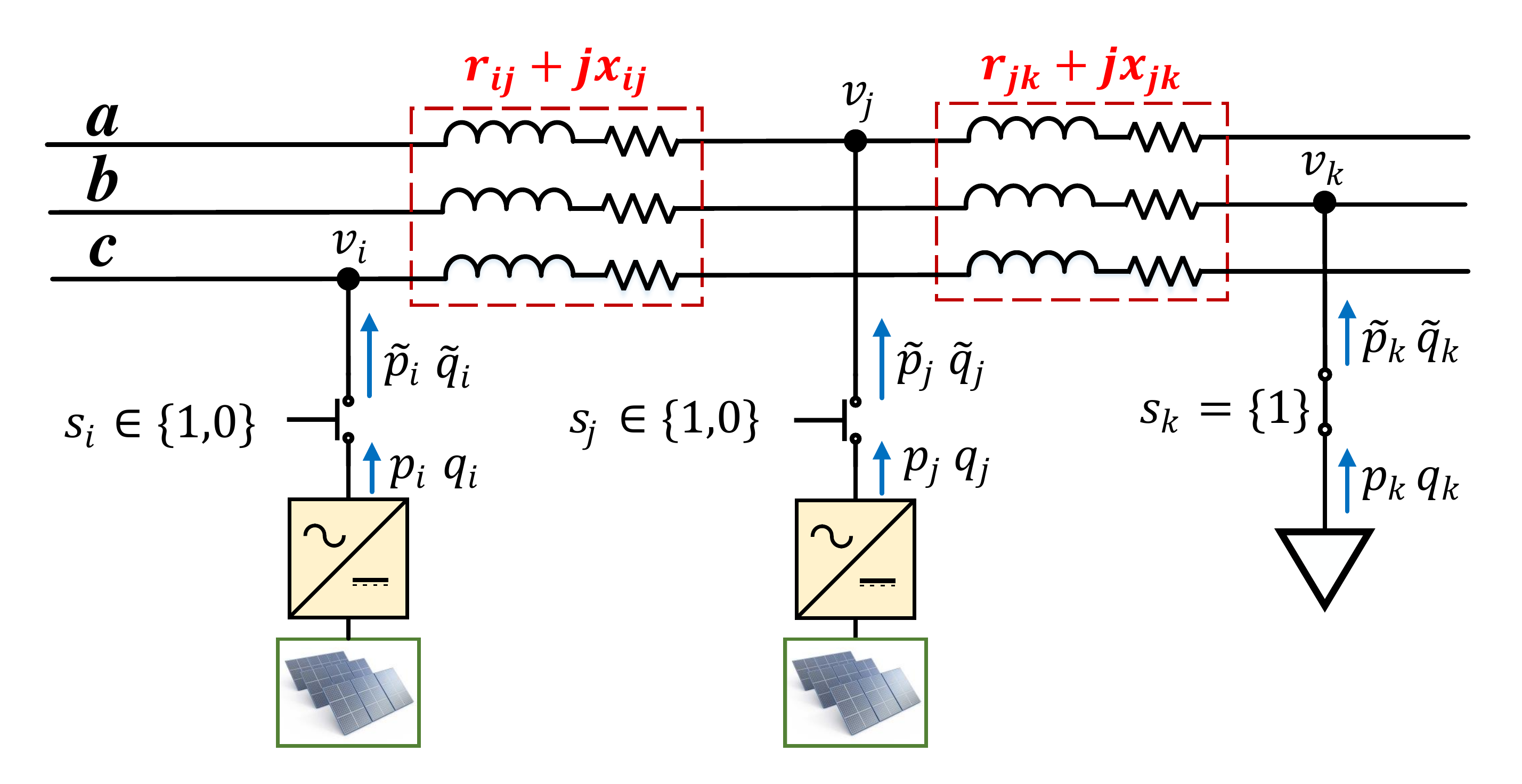}
\caption{Distribution feeder structure with PVs, loads, and voltage-sensitive current interruption mechanisms (i.e., switches).}
\label{fig:main}
\end{figure}
\subsection{Original Interactive Switching Equations}
In this paper, it is assumed that PV resources are protected against voltage deviations using ON/OFF switching mechanisms. Note that here a ``switch'' can be a mechanical relay, as well as a \textit{non-physical} inverter control function that stops current injection into the grid under abnormal voltage even if the inverter is still physically connected to the grid \cite{wecc}. The PV is tripped in case the nodal voltage deviates from a user-defined permissible range, $[V_{min},V_{max}]$. In this paper, this range is adopted from the literature \cite{Ferreira2013}, as $V_{min} = 0.9\ p.u.$ and $V_{max} = 1.1\ p.u.$. The switching mechanisms are simply modelled as binary \textit{micro-state} variables with the following voltage-dependent function (see Fig. \ref{fig:main}):

\begin{equation}
\label{sw}
s_i(t) = 
\begin{cases}
1\ \ \ \ V_{min}\leq V_i(t) \leq V_{max}\\
0\ \ \ \ V_i(t) < V_{min}\\
0\ \ \ \ V_i(t) > V_{max}
\end{cases}
\end{equation}
where, $s_i(t)$ is the micro-state assigned to the $i$'th PV at time $t$ as a function of the inverter node's voltage magnitude $V_i$. Here, $s_i(t) = 1$ implies ON and $s_i(t) = 0$ indicates OFF. The assumption in this switching model is that over long enough time intervals the impact of inverter dynamics, e.g., ride-through capabilities \cite{NERC, AEMO2018, AEMO2017,Dozein2018}, can be conservatively ignored. This assumption considerably enhances the tractability of the model at the expense of loss of accuracy. In this sense, the switching model is a worst-case representation of inverter tripping. Since the approximate power flow equations for distribution grids are linear with respect to the squared values of nodal voltage magnitudes \cite{Li2014}, we re-write equation \eqref{sw} using a variable transformation, $v_i = V_i^2$, and employing unit step functions as follows:
\begin{equation}
\label{eq:switch}
s_i(t) = U(v_i(t) - v_{min}) - U(v_i(t) - v_{max})
\end{equation}
where, $v_{min} = V_{min}^2$, $v_{max} = V_{max}^2$, and the unit step function $U(\cdot)$ is defined as follows:
\begin{equation}
U(x) = 
\begin{cases}
1\ \ \ \ x \geq 0\\
0\ \ \ \ x < 0,
\end{cases}
\end{equation}

Note that inverters' micro-states are influenced by nodal voltages and are thus highly interdependent on each other, as changes in the state of one switch will cause nodal power variations, which leads to a change of voltage at other nodes that can in turn influence probability of tripping events. To obtain the overall governing equations of inverter tripping, the mutual impacts of switch micro-states on each other are captured using an approximate unbalanced power flow model for radial distribution grids \cite{Li2014}, which determines voltage at node $i$ as a function of active/reactive power injections of every other node in a grid (with a total of $N+1$ nodes):
\begin{equation}
\label{eq:vi_vij}
v_i(t) = \sum_{j=1}^N\Tilde{v}_{ij} + v_0,\ \forall i\in\{1,...,N\}
\end{equation}
where, $v_0 = V_0^2$, with $V_0$ denoting the voltage magnitude at a grid reference bus, and the intermediary variable $\Tilde{v}_{ij}$ represents the impact of active/reactive power injection at node $j$ on $v_i$, which is obtained as follows:  
\begin{equation}
\label{eq:vijpjqj}
\Tilde{v}_{ij} = R_{ij}\Tilde{p}_j(t) +  X_{ij}\Tilde{q}_j(t)
\end{equation}
where, $R_{ij}$ and $X_{ij}$ are the aggregated series resistance and reactance values corresponding to the intersecting branches in the paths connecting nodes $i$ and $j$ to the reference bus calculated as follows \cite{Li2014}: 
\begin{equation}
\label{eq:Rij}
R_{ij} = 2\sum_{\{n,m\}\in Pa(i,j)}r_{nm}
\end{equation}
\begin{equation}
\label{eq:Xij}
X_{ij} = 2\sum_{\{n,m\}\in Pa(i,j)}x_{nm}
\end{equation}
where, $Pa(i,j)$ represents the set of pairwise nodes consisting of the neighboring nodes that are on the intersection of the unique paths connecting nodes $i$ and $j$ to the reference bus; $r_{nm}$ and $x_{nm}$ denote the real series resistance and reactance of the branch connecting nodes $n$ and $m$. Also, $\Tilde{p}_j$ and $\Tilde{q}_j$ denote the active and reactive power injections at bus $j$, which are in turn determined by the micro-state of the PV at node $j$ (see Fig. \ref{fig:main}): 
\begin{equation}
\label{eq:ptilde}
\Tilde{p}_j(t) = p_j(t)s_j(t)
\end{equation}
\begin{equation}
\label{eq:qtilde}
\Tilde{q}_j(t) = q_j(t)s_j(t)
\end{equation}
with $p_j$ and $q_j$ representing the available load/PV power at node $j$, where $p_j > 0$ implies generation. Equations \eqref{eq:vi_vij}-\eqref{eq:Xij} are obtained in vector form for all three phases of unbalanced distribution grids \cite{Li2014}. 

Equations \eqref{eq:switch}-\eqref{eq:qtilde} fully determine the states of networked PVs. The difficulty in solving these equations is due to three factors: (I) the size of solution space increases exponentially as the number of micro-states $\{s_1,...,s_N\}$ grows. Since these micro-states are not independent and influence each other in complex and non-trivial ways they cannot be obtained individually, and a thorough search process is needed to explore all possible switching configurations. This can be extremely expensive and impossible to scale to large systems with high population of inverters. (II) Due to the discrete step functions in \eqref{eq:switch}, tripping equations are nonlinear and non-differentiable. This contributes to problem difficulty since gradient-based methods cannot be applied. (III) $p_j$ and $q_j$ act as time-varying input parameters within the model. This implies that using the tripping equations for predicting probability of tripping events requires extensive search process to cover all probable PV/load time-series scenarios. This expensive search process hinders the tractability of optimization-based frameworks for designing tripping mitigation strategy.

Not all the nodes in the tripping model are necessarily controlled by ON/OFF voltage-sensitive switching mechanisms. For examples, ordinary load nodes are generally not governed by equation \eqref{eq:switch}. In this paper, for the sake of brevity, the switching equations are still written for all the nodes in the grid as presented, however, we will simply assign constant values, $s_i(t) = 1,\ \forall t$ to the nodes without ON/OFF control and remove their corresponding switching from the equations (see Fig. \ref{fig:main}). 

\subsection{Alternative Approximate Probabilistic Model}
We adopt a probabilistic point of view towards tripping model. This allows us to obtain a stationary differentiable statistical model that has a simple matrix-form formulation. Accordingly, the ON/OFF current interruption mechanisms, $s_i$'s, are modelled as random variables following Bernoulli probability distributions with parameters $\lambda_i, \forall i \in \{1,...,N\}$: $s_i\sim\mathpzc{B}(\lambda_i)$, where parameter $\lambda_i$ is defined as the probability of the $i$'th inverter switch being ON, $\lambda_i(t) = Pr\{s_i(t) = 1\}$. The goal is to transform micro-states from discontinuous binary variables ($s_i \in \{0,1\}$) into continuous variables ($\lambda_i \in [0,1]$). To rewrite the equations in terms of new micro-states note that we have $E\{s_i(t)\} = \lambda_i(t)$ for Bernoulli probability distributions, where $E\{\cdot\}$ represents the expectation operation. Thus, by performing an expectation operation over both sides of \eqref{eq:switch}, probability of inverter tripping in terms of the new micro-states can be obtained as follows: 
\begin{equation}
\label{eq:1switchAve}
\lambda_i(t) = Pr\{v_{min} \leq v_i(t) \leq v_{max}\}
\end{equation}
where, we have exploited $E\{U(f(x))\} = Pr\{f(x)\geq 0\}$. Note that the probability of tripping for an inverter is an implicit function of nodal voltage probability distribution, which in turn is influenced by the states of other inverters. Due to the interconnected nature of the problem, no independency assumptions has been made on random variables $\lambda_i, \forall i \in \{1,...,N\}$. However, the exact distributions of nodal voltages are unknown and complex functions of nodal active/reactive injections, which implies that \eqref{eq:1switchAve} cannot be determined analytically unless over-simplifying assumptions are made. Instead, we employ Chebyshev's inequality \cite{Ross2006} to provide a lower bound on micro-state as a function of nodal voltage statistics without making any assumption on voltage distributions, 
\begin{equation}
\label{eq:Cheb}
Pr\{v_{min} \leq v_i(t) \leq v_{max}\} \geq 1 - \frac{\sigma_{v_i}^2 + (\mu_{v_i} -\frac{v_{max} + v_{min}}{2})^2}{(\frac{v_{max} - v_{min}}{2})^2}
\end{equation}
where, $\sigma_{v_i}^2$ and $\mu_{v_i}$ are the variance and mean of $v_i$, respectively. Hence, the approximate probabilistic model can be formulated for each micro-state as follows:  
\begin{equation}
\label{eq:approx}
\hat{\lambda}_i(t) = 1 - \frac{\sigma_{v_i}^2 + (\mu_{v_i} -\frac{v_{max} + v_{min}}{2})^2}{(\frac{v_{max} - v_{min}}{2})^2}
\end{equation}

This new tripping model has two features: (1) it is a conservative estimator of the original system since it over-estimates the probability of inverter tripping, $\hat{\lambda}_i \leq \lambda_i$. (2) As will be shown in Section \ref{sec:para}, the approximate probabilistic model can be conveniently parameterized in terms of nodal available active/reactive power statistics. Hence, as long as certain statistics are known (or estimated), the model allows us to accurately track probability of inverter tripping without running time-series simulations under numerous scenarios.

\subsection{Probabilistic Model Parameterization}\label{sec:para}
To parameterize the alternative tripping model \eqref{eq:approx}, nodal voltage statistics, $\sigma_{v_i}^2$ and $\mu_{v_i}$, are obtained in terms of nodal available active/reactive power statistics. To do this, power flow/injection equations \eqref{eq:vi_vij}-\eqref{eq:qtilde} are leveraged. 

\textit{\textbf{Stage 1:} \textbf{$\mu_{v_i}$ Parameterization}} - The expected value of voltage magnitude squared is determined using \eqref{eq:vi_vij}-\eqref{eq:vijpjqj} as,
\begin{align}
\label{eq:Evij}
\mu_{v_i} =&  \sum_{j=1}^NE\{\Tilde{v}_{ij}\} + v_0 \nonumber\\
=&  \sum_{j=1}^N(R_{ij}E\{\Tilde{p}_j\} +  X_{ij}E\{\Tilde{q}_j\}) + v_0
\end{align}

To calculate $E\{\Tilde{p}_j\}$ and $E\{\Tilde{q}_j\}$, we will first obtain their cumulative distribution functions (CDFs) \cite{Ross2006}, $F_{\tilde{p}_j}$ and $F_{\tilde{q}_j}$, respectively. This process is shown for $\Tilde{p}_j$ as follows ($F_{\tilde{q}_j}$ is obtained similarly): 
\begin{equation}
\label{eq:ptildeCDF}
F_{\tilde{p}_j}(P) = Pr\{\Tilde{p}_j(t)\leq P\} = (1 - \lambda_j(t))U(P) + \lambda_j(t)F_{p_j}(P)
\end{equation}

The rational behind \eqref{eq:ptildeCDF} is that the distribution of power injection is determined by two functions: the distribution of PV switch (which is ON with probability $\lambda_j(t)$), and the CDF of available PV power, $F_{p_j}$. Now, the probability density functions (PDF) of the realized active nodal power injection, $f_{\tilde{p}_j}$, can be calculated as a function of the available active solar power, $f_{p_j}$ (a similar operation is performed for reactive power):
\begin{equation}
\label{eq:ptildePDF}
f_{\tilde{p}_j}(P) = \dv{F_{\tilde{p}_j}(P)}{P} = (1 - \lambda_j(t))\delta(P) + \lambda_j(t)f_{p_j}(P)
\end{equation}

Then, using the active/reactive power injection PDFs, $E\{\Tilde{p}_j\}$ and $E\{\Tilde{q}_j\}$, can be obtained through integration:
\begin{equation}
\label{eq:ptildeExpP}
E\{\tilde{p}_j\} = \int_{-\infty}^{+\infty}\alpha f_{\tilde{p}_j}(\alpha) \mathrm{d}\alpha = \lambda_j P_j
\end{equation}
\begin{equation}
\label{eq:ptildeExpQ}
E\{\tilde{q}_j\} = \int_{-\infty}^{+\infty}\beta f_{\tilde{q}_j}(\beta) \mathrm{d}\beta = \lambda_j Q_j
\end{equation}
where, $P_j$ and $Q_j$ denote the mean values of the available active and reactive powers at node $j$, respectively ($P_j = E\{p_j\}$ and $Q_j = E\{q_j\}$). Thus, the mean nodal voltage magnitude squared can be written in terms of inverter switch statistics and expected PV/load available powers: 
\begin{equation}
\label{eq:mu}
\mu_{v_i} = \sum_{j=1}^{N}\{R_{ij}\lambda_j(t)P_j + X_{ij}\lambda_j(t)Q_j\} + v_0
\end{equation}

\textit{\textbf{Stage 2: $\sigma_{v_i}^2$ Parameterization}} - Using \eqref{eq:vi_vij}, the variance of nodal voltage magnitude squared can be formulated as,
\begin{equation}
\label{eq:sigv}
\sigma_{v_i}^2 = \sum_{j=1}^{N}\sigma_{\Tilde{v}_{ij}}^2 + 2 \sum_{1\leq k<j\leq N}\Omega\{\Tilde{v}_{ij},\Tilde{v}_{ik}\}
\end{equation}
where, $\sigma_{\Tilde{v}_{ij}}^2$ is the variance of $\Tilde{v}_{ij}$, and the operator $\Omega\{x_1,x_2\}$ denotes the covariance of the two random variables $x_1$ and $x_2$, which itself can be written in terms of their correlation, $\rho_{x_1,x_2}$, and standard deviations, $\sigma_{x_1}$ and $\sigma_{x_2}$, as  $\Omega\{x_1,x_2\} = \rho_{x_1,x_2}\sigma_{x_1}\sigma_{x_2}$. To fully parameterize $\sigma_{v_i}^2$ using available load/PV power statistics, $\sigma_{\Tilde{v}_{ij}}^2$ and $\Omega\{\Tilde{v}_{ij},\Tilde{v}_{ik}\}$ have to be determined separately.

\textit{\textbf{Stage 2-1: $\sigma_{\Tilde{v}_{ij}}^2$ Parameterization}} - Using \eqref{eq:vijpjqj}, $\sigma_{\Tilde{v}_{ij}}^2$ is formulated as a function of $\tilde{p}_j$ and $\tilde{q}_j$ statistics: 
\begin{equation}
\label{eq:sigvtilde}
\sigma_{\Tilde{v}_{ij}}^2 = R_{ij}^2\sigma_{\tilde{p}_j}^2 + X_{ij}^2\sigma_{\tilde{q}_j}^2 + 2R_{ij}X_{ij}\Omega\{\tilde{p}_j,\tilde{q}_j\}
\end{equation}
where, $\sigma_{\tilde{p}_j}^2$ and $\sigma_{\tilde{q}_j}^2$ are the active/reactive power injection variances, which can in turn be determined as follows:

\begin{equation}
\label{eq:tildesq}
\sigma_{\Tilde{p}_{j}}^2 = E\{s_j^2p_j^2\} - E\{\tilde{p}_j\}^2
\end{equation}
where, $E\{s_j^2p_j^2\}$ is calculated through a similar process involved in \eqref{eq:ptildeCDF}-\eqref{eq:ptildeExpQ} (i.e., obtain the CDF, determine the PDF, and integrate). Noting that in our case $s_j^2 = s_j$, the PDF of $s_j^2p_j^2$ is derived as follows (similar derivation applies to $s_j^2q_j^2$):
\begin{equation}
\label{eq:pdfPssqq2}
f_{s_j^2p_j^2}(\zeta) = (1 - \lambda_j(t))\delta(\zeta) + \frac{\lambda_j(t)}{2\sqrt{\zeta}}(f_{p_j}(\sqrt{\zeta}) + f_{p_j}(-\sqrt{\zeta}))
\end{equation}

By integrating \eqref{eq:pdfPssqq2} and using \eqref{eq:ptildeExpP}-\eqref{eq:ptildeExpQ} to substitute for $E\{\tilde{p}_j\}$ and $E\{\tilde{q}_j\}$, the following results are obtained to parameterize the variances of nodal active/reactive power injections:
\begin{equation}
\label{eq:varsqP}
\sigma_{\Tilde{p}_{j}}^2 = \lambda_j(P_j^+ + P_j^-) - \lambda_j^2P_j^2
\end{equation}
\begin{equation}
\label{eq:varsqQ}
\sigma_{\Tilde{q}_{j}}^2 = \lambda_j(Q_j^+ + Q_j^-) - \lambda_j^2Q_j^2
\end{equation}
where, $P_j^+ = E\{p_j^2|p_j\geq0\}$ and $P_j^- = E\{p_j^2|p_j<0\}$; similar definitions apply to $Q_j^+$ and $Q_j^-$. Note that given that $p_j\geq0$ for PVs, $P_j^+ = \sigma_{p_j}^2 + P_j^2$ and $P_j^- = 0$. Employing an analogous logic, $P_j^+ = 0$ and $P_j^- = \sigma_{p_j}^2 + P_j^2$ for loads.   

To obtain $\Omega\{\tilde{p}_j,\tilde{q}_j\}$ in \eqref{eq:sigvtilde}, we leverage the fact that $\Omega\{x_1,x_2\} = E\{x_1x_2\} - E\{x_1\}E\{x_2\}$ as follows:
\begin{equation}
\label{eq:covjj}
\Omega\{\tilde{p}_j,\tilde{q}_j\} = E\{\tilde{p}_j\tilde{q}_j\} - E\{\tilde{p}_j\}E\{\tilde{q}_j\}
\end{equation}
where, the term $E\{\tilde{p}_j\tilde{q}_j\}$ is calculated similar to previous derivations (i.e., CDF$\rightarrow$PDF$\rightarrow$integration), which combined with \eqref{eq:ptildeExpP} and \eqref{eq:ptildeExpQ} yields the following result:
\begin{equation}
\label{eq:covjjfinal}
\Omega\{\tilde{p}_j,\tilde{q}_j\} = \lambda_jP_jQ_j - \lambda_j^2P_jQ_j + \lambda_j\Omega\{p_j,q_j\}
\end{equation}
where, $\Omega\{p_j,q_j\}$ can be determined in terms of available active/reactive power statistics, including correlation and standard deviations as $\Omega\{p_j,q_j\} = \rho_{p_j,q_j}\sigma_{p_j}\sigma_{q_j}$.

Thus, using \eqref{eq:varsqP}, \eqref{eq:varsqQ}, and \eqref{eq:covjjfinal}, $\sigma_{\Tilde{v}_{ij}}^2$ can be parameterized in terms of the available active/reactive power statistics, and with respect to micro-states:
\begin{equation}
\label{eq:sigvtildefinal}
\sigma_{\Tilde{v}_{ij}}^2 = \lambda_j\Gamma_{ij}^1 -\lambda_j^2\Gamma_{ij}^2
\end{equation}
where, the time-invariant parameters $\Gamma_{ij}^1$ and $\Gamma_{ij}^2$ are given below:
\begin{align}
\label{eq:gamma1ij}
\Gamma_{ij}^1 = &R_{ij}^2(P_j^+ + P_j^-) + X_{ij}^2(Q_j^+ + Q_j^-)\nonumber\\
&+ 2R_{ij}X_{ij}(P_jQ_j + \Omega\{P_j,Q_j\})
\end{align}
\begin{equation}
\label{eq:gamma1ij}
\Gamma_{ij}^2 = 2R_{ij}X_{ij}P_jQ_j + P_j^2R_{ij}^2 + Q_j^2X_{ij}^2
\end{equation}

\textit{\textbf{Stage 2-2: $\Omega\{\tilde{v}_{ij},\tilde{v}_{ik}\}$ Parameterization}} - Similar to \eqref{eq:covjjfinal}, $\Omega\{\tilde{v}_{ij},\tilde{v}_{ik}\}$, is broken down to its components: 
\begin{equation}
\label{eq:covvijvik1}
\Omega\{\tilde{v}_{ij},\tilde{v}_{ik}\} = E\{\tilde{v}_{ij}\tilde{v}_{ik}\} - E\{\tilde{v}_{ij}\}E\{\tilde{v}_{ik}\}
\end{equation}
By adopting a CDF$\rightarrow$PDF$\rightarrow$integration strategy, $E\{\tilde{v}_{ij}\tilde{v}_{ik}\}$ is determined in terms of active/reactive power injection statistics as follows:
\begin{align}
\label{eq:covvijvik2}
E\{\tilde{v}_{ij}\tilde{v}_{ik}\} = &R_{ij}R_{ik}E\{\tilde{p}_j,\tilde{p}_k\} + R_{ij}X_{ik}E\{\tilde{p}_j,\tilde{q}_k\} +\nonumber\\ &X_{ij}R_{ik}E\{\tilde{q}_j,\tilde{p}_k\} + X_{ij}X_{ik}E\{\tilde{q}_j,\tilde{q}_k\}
\end{align}
where, using previous derivations and through algebraic manipulations, the following parameterization is obtained in terms of available active/reactive power statistics for $\Omega\{\tilde{v}_{ij},\tilde{v}_{ik}\}$:
\begin{equation}
\label{eq:covvijvikF}
\Omega\{\tilde{v}_{ij},\tilde{v}_{ik}\} = \lambda_j\lambda_k (\Gamma_{ijk}^1 - \Gamma_{ijk}^2)
\end{equation}
where, the parameters $\Gamma_{ijk}^1$ and $\Gamma_{ijk}^2$ are determined as:
\begin{align}
\label{eq:gammaijk1}
&\Gamma_{ijk}^1 = \nonumber\\
&R_{ij}R_{ik}(\Omega\{p_j,p_k\} + P_jP_k) + R_{ij}X_{ik}(\Omega\{p_j,q_k\} + P_jQ_k) +\nonumber\\ &X_{ij}R_{ik}(\Omega\{q_j,p_k\} + Q_jP_k) + X_{ij}X_{ik}(\Omega\{q_j,q_k\} + Q_jQ_k)
\end{align}
\begin{equation}
\label{eq:gammaijk2}
\Gamma_{ijk}^2 = (R_{ij}P_j + X_{ij}Q_j)(R_{ik}P_k + X_{ik}Q_k)
\end{equation}

By substituting \eqref{eq:covvijvikF} and \eqref{eq:sigvtildefinal} into \eqref{eq:sigv}, $\sigma_{v_i}^2$ is now fully determined as a function of micro-states and in terms of available nodal active/reactive power statistics.

\textit{\textbf{Stage 3. Probabilistic Inverter Tripping Model Representation:}} Finally, using the parameterized $\sigma_{v_i}^2$ and $\mu_{v_i}$, the probabilistic model \eqref{eq:approx} yields a the following bilinear matrix-form representation for the approximate micro-state vector $\pmb{\hat{\lambda}} = [\hat{\lambda}_1,...,\hat{\lambda}_N]^\top$:  
\begin{equation}
\label{eq:microApprox}
\pmb{\hat{\lambda}}(t) = \pmb{a_0} + B\pmb{\hat{\lambda}}(t) + \left[
\begin{array}{c}
\pmb{\hat{\lambda}}(t)^\top C_1\pmb{\hat{\lambda}}(t) \\
\vdots\\
\pmb{\hat{\lambda}}(t)^\top C_N\pmb{\hat{\lambda}}(t)
\end{array}
\right]
\end{equation}
where, all the time-invariant parameters of the model are concatenated into the vector $\pmb{a_0}$ and matrices $B$, and $\{C_1, ..., C_N\}$. The elements of these parameters are determined by organizing the previous derivations in Stages 1 and 2, as follows: 
\begin{equation}
\label{eq:a0}
\pmb{a_0}(i) = 1 - (\frac{2v_0 - v_{max} - v_{min}}{v_{max} - v_{min}})^2
\end{equation}
\begin{align}
\label{eq:B}
&B(i,j) = \nonumber\\
&\frac{-1}{(\frac{v_{max} - v_{min}}{2})^2}\Gamma_{ij}^1 - \frac{2v_0 - v_{max} - v_{min}}{(\frac{v_{max} - v_{min}}{2})^2}(P_jR_{ij}+Q_jX_{ij})
\end{align}
\begin{equation}
\label{eq:C}
C_i(j,k) = 
\begin{cases}
\frac{-1}{(\frac{v_{max} - v_{min}}{2})^2}\Gamma_{ijk}^1\ \ \ \ j \neq k\\
0\ \ \ \ \ \ \ \ \ \ \ \ \ \ \ \ \ \ \ \ \ \ \ j = k,
\end{cases}
\end{equation}
where, $\pmb{a_0}(i)$ denotes the $i$'th element of $\pmb{a_0}$, and $B(i,j)$ and $C_i(j,k)$ are the $(i,j)$'th and $(j,k)$'th elements of $B$ and $C_i$, respectively. The aggregate switching equation can be written as a function of approximate \textit{macro-state}, $\hat{S} = \sum_{i=1}^N\hat{\lambda}_i$, as follows: 
\begin{equation}
\label{eq:macroApprox}
\hat{S}(t) = [\sum_{i=1}^{N}\pmb{a_0}(i)] + [\sum_{i=1}^{N} B(i,:)]\cdot\pmb{\hat{\lambda}}(t) +
\pmb{\hat{\lambda}}(t)^\top [\sum_{i=1}^{N} C_i]\pmb{\hat{\lambda}}(t)
\end{equation}
where, $\hat{S}$ is a conservative estimator of the real macro-state, $S$, which is the actual expected population of inverter that are ON, i.e., $\hat{S}(t) \leq S(t)$. Also, $B(i,:)$ is the $i$'th row of matrix $B$. To summarize, the proposed approximate probabilistic model leverages available load/PV power statistics shown in Table \ref{table:stat}. Previous works have used various data-driven and machine learning methods that can be applied for obtaining statistical properties of nodal load/PV powers in partially observable networks from limited available data (for example see \cite{Singh2010,Nguyen2015,Hatton2016}). Also, although the micro-states in the probabilistic model are random variables, the model itself is governed by deterministic functions of load/PV statistics.

A related problem in distribution grids that testifies to the dependent nature of tripping is known as \textit{sympathetic tripping} of inverters in weak grids \cite{Jennett2014, Walt2018}: overloading/faults on one feeder can trigger the voltage protection mechanism of inverters on a healthy neighboring feeder. The sympathetic tripping of inverters is also caused by dependencies in nodal voltages within the distribution grid (i.e., excessive load/fault current on one feeder contributing to voltage drops on other nodes). While sympathetic tripping is not exactly what the proposed statistical model in this paper captures, it still provides further support that dependency in tripping is possible in practice.

\begin{table}
\begin{center}
 \caption{Needed Statistics for Developing the Proposed Model}
  \includegraphics[width=0.9\linewidth]{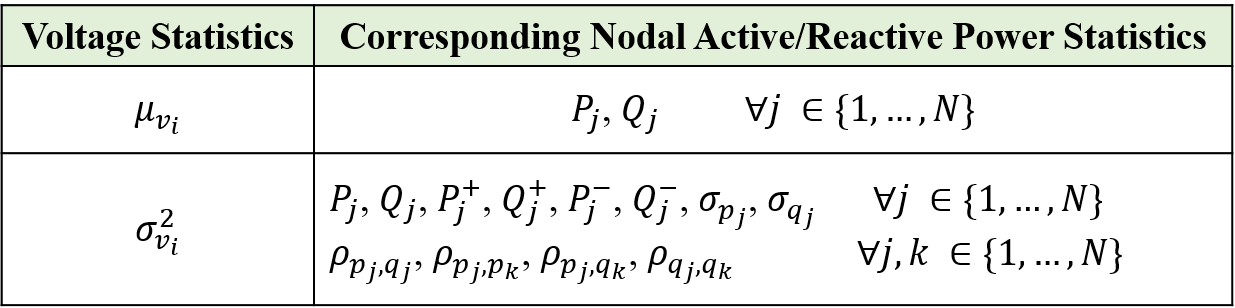}
    \label{table:stat}
\end{center}
 \vspace{-1.25em}
\end{table}

\subsection{Discussion on Probabilistic Tripping Model Properties}\label{sec:prop}
The probabilistic model \eqref{eq:microApprox} represents a set of \textit{self-consistent} equations; in other words, any $\pmb{\hat{\lambda}}$ that satisfies these equations is a conservative estimator of probability of inverter tripping. Furthermore, this probabilistic model can be thought of as the asymptotic equilibrium of an \textit{abstract discrete dynamic system}:
\begin{equation}
\label{eq:abstractdyn}
\pmb{\hat{\lambda}}(k+1) = \pmb{a_0} + B\pmb{\hat{\lambda}}(k) + \left[
\begin{array}{c}
\pmb{\hat{\lambda}}(k)^\top C_1\pmb{\hat{\lambda}}(k) \\
\vdots\\
\pmb{\hat{\lambda}}(k)^\top C_N\pmb{\hat{\lambda}}(k)
\end{array}
\right]
\end{equation}
where, the equilibrium is achieved at $\pmb{\hat{\lambda}}(k+1) = \pmb{\hat{\lambda}}(k)$ and coincides with the solution of the proposed probabilistic model \eqref{eq:microApprox}. This abstract dynamic system has an intuitive interpretation: matrix $B$ represents the \textit{linear} component of the dynamics, which as can be observed in \eqref{eq:B} and \eqref{eq:gamma1ij}, is determined only by each individual nodes' active/reactive power statistics, including the expected values and self-correlation between active/reactive power at each node alone. However, matrices $\{C_1,...,C_N\}$ capture the \textit{nonlinear} components of the dynamic system, where the element $C_i(j,k)$ determines the coefficient assigned to the interactive nonlinear probability-product term $\hat{\lambda_j}(t)\cdot\hat{\lambda}_k(t)$ in driving $\hat{\lambda_i}(t+1)$. In other words, $C_i(j,k)$ quantifies the mutual impact of the $j$'th and $k$'th PV micro-states on dynamics of the $i$'th switch. Furthermore, as observed in \eqref{eq:C} and \eqref{eq:gammaijk1} the elements of $C_i$, unlike $B$, are determined by the mutual correlations in available active/reactive powers of different PVs. The inherent nonlinearity of \eqref{eq:abstractdyn} hints at the possibility of \textit{stage transition} and bifurcation at equilibrium of the abstract dynamic system as PV/load power statistics evolve over time, which could potentially result into a cascading tripping event, as pointed out in \cite{Ferreira2013,Popiel,AEMO2018,AEMO2017}. A regime shift at the equilibrium of the abstract nonlinear dynamic system basically corresponds to qualitative changes in the solution of our probabilistic tripping model, potentially, leading to a sudden increase in the average chances of voltage-driven tripping events caused by the growing penetration of solar energy in the system. In this sense, the structure of the abstract dynamic model is similar to other complex interactive dynamic systems in the literature, including nonlinear combinatorial evolution models \cite{Thurner2018} and asymmetric Ising systems \cite{Mezard2011}, which are also known to demonstrate critical behavior and emergent non-trivial patterns at the macro-level under certain conditions.

An important factor in tripping studies is the impact of setting of inverter protection systems. This can be seen in equations \eqref{eq:a0}, \eqref{eq:B}, and \eqref{eq:C} that present the parameters of the proposed statistical model. Specifically, parameter $C_i(j,k)$ in \eqref{eq:C}, captures the joint impacts of inverter $j$ and inverter $k$ on probability of tripping for inverter $i$. As can be seen, the absolute value of this parameter decreases with $\frac{1}{(v_{max} - v_{min}) ^2}$. Thus, increasing the inverter upper protection threshold, $v_{max}$, or decreasing the lower protection threshold, $v_{min}$, (i.e., making the inverter less sensitive to voltage events) will result in a decline in mutual impacts of inverters on each other. In other words, relaxing the protection boundaries significantly weakens the dependencies among inverter tripping. If $C_i(j,k)$ is thought of as a measure of strength of interdependency among inverters, then our model suggests that loss of interdependency is approximately proportional to the inverse of inverter protection dead-band width squared.

\subsection{Integrating Voltage-Dependent Resources Into the Proposed Probabilistic Tripping  Model}\label{sec:voltdep}
Note that so far we have assumed that the nodal active and reactive power injections, $p_j$ and $q_j$, are external inputs to the model. However, active and reactive power injection of certain nodes can show high levels of voltage-dependency and cannot be treated as external inputs. The voltage-dependency can be caused by reactive power support from the inverters or load power voltage-sensitivity. In this section, we will demonstrate that voltage-dependent resources can also be included in our probabilistic model. To do this, the active/reactive power injections are linearized around the nominal squared voltage ($v_n$):
\begin{equation}
\label{eq:pdelta}
p_j(v_j)\approx p_j(v_n) + {\left.\dv{p_j(v_j)}{v_j}\right|_{v_j=v_n}}\times(v_j-v_n)
\end{equation}
\begin{equation}
\label{eq:qdelta}
q_j(v_j)\approx q_j(v_n) + {\left.\dv{q_j(v_j)}{v_j}\right|_{v_j=v_n}}\times(v_j-v_n)
\end{equation}
The active/reactive power injections in \eqref{eq:pdelta} and \eqref{eq:qdelta} consist of two terms: one is the voltage-independent term, and the second is caused by non-zero sensitivity to nodal voltage. Our model can conveniently include the first term as outlined previously. The second term can also be integrated in the model if the operator has a rough estimation of active/reactive power voltage-sensitivity values. For example, this sensitivity can be obtained for ZIP loads \cite{anmar} and inverters that are capable of reactive power support \cite{AEMO2018,AEMO2017} as follows:
\begin{equation}
\label{eq:psens}
{\left.\dv{p_j(v_j)}{v_j}\right|_{v_j=v_n}} = p_j(v_n)\cdot(\frac{B_j+2C_j}{2v_n})
\end{equation}
\begin{equation}
\label{eq:qsens}
{\left.\dv{q_j(v_j)}{v_j}\right|_{v_j=v_n}} = k_j
\end{equation}
where, $B_j$ and $C_j$ represent the ZIP coefficients corresponding to the fixed-current and fixed-impedance portions of ZIP load, respectively, and $k_j < 0$ is the local inverter droop coefficient. Given the voltage-sensitivity values, the second terms in \eqref{eq:pdelta} and \eqref{eq:qdelta} simply serve as new additional nodal active/reactive power injections and can be treated in the model similar to other loads. For example, the surrogate nodal active/reactive injections for ZIP loads and inverters with reactive support capability can be conservatively estimated as follows:
\begin{equation}
\label{eq:pnew}
\Delta p_j \approx (\frac{B_j+2C_j}{2v_n})(\Bar{v} - v_n)p_j(v_n)
\end{equation}
\begin{equation}
\label{eq:qnew}
\Delta q_j \approx k_j(\Bar{v} - v_n)
\end{equation}
where, $\Bar{v}$ denotes a conservative user-defined value that can be used by the utilities to model worst-case tripping scenarios. However, note that \eqref{eq:pnew} and \eqref{eq:qnew} are still conservative estimations. Developing more accurate models for integrating voltage-dependent power injection into tripping equations remains the subject of future research.  

\section{Solar Curtailment Quantification and Mitigation}\label{sec:opt}
Using \eqref{eq:microApprox} as a conservative probabilistic lower bound for the real system, an optimization problem is formulated to provide a realistic estimation of the actual values of the micro-states of the grid. This problem is solved at any given time-window at which available nodal active/reactive power statistics are known:
\begin{equation}
\label{eq:Opt}
\begin{split}
&\min_{\pmb{\hat{\lambda}}} -(\pmb{P}^\top\cdot\pmb{\hat{\lambda}}),\\
 s.t.\ \ \ \pmb{\hat{\lambda}} = \pmb{a_0}& + B\pmb{\hat{\lambda}} + \left[
\begin{array}{c}
\pmb{\hat{\lambda}}^\top C_1\pmb{\hat{\lambda}} \\
\vdots\\
\pmb{\hat{\lambda}}^\top C_N\pmb{\hat{\lambda}}
\end{array}
\right]\\
0 \leq \hat{\lambda}_j& \leq 1\ \ \forall j\in\{1,...,N\}
\end{split}
\end{equation}
where, $\pmb{P} = [P_1,...,P_N]^\top$. The objective of this optimization problem is to find the maximum achievable expected solar power in the gird according to the conservative statistical model. While the solution to this problem is still a lower bound estimation of the real achievable PV power, the estimation gap between $\pmb{\hat{\lambda}}$ and $\pmb{\lambda}$ is minimized. In other words, the optimization searches for the most optimistic values for micro-states with respect to the conservative approximate probabilistic tripping model. The problem is constrained by the matrix equations that govern the probabilities of inverter tripping. Furthermore, the physical characteristics of micro-states are constrained by valid probability assignments within $[0,1]$ interval.

A similar problem can be formulated to provide counter-measures against massive tripping events at any given time window. In general, the proposed statistical tripping model can be integrated as a constraint into any volt-var optimization formulation \cite{Petinrin2016,Ranamuka2014,Ku2019} to represent the possibility of PV curtailment. For example, here we provide a formulation for minimizing solar curtailment by controlling the voltage magnitude at the system reference bus \cite{Petinrin2016}:
\begin{equation}
\label{eq:OptControl}
\begin{split}
&\min_{\pmb{\hat{\lambda}},v_0} -(\pmb{P}^\top\cdot\pmb{\hat{\lambda}}),\\
 s.t.\ \ \ \pmb{\hat{\lambda}} = \pmb{a_0}&(v_0) + B(v_0)\pmb{\hat{\lambda}} + \left[
\begin{array}{c}
\pmb{\hat{\lambda}}^\top C_1\pmb{\hat{\lambda}} \\
\vdots\\
\pmb{\hat{\lambda}}^\top C_N\pmb{\hat{\lambda}}
\end{array}
\right]\\
0 \leq \tilde{\lambda}_j& \leq 1\ \ \forall j\in\{1,...,N\}\\
&v_{min} \leq v_0 \leq v_{max}\\
v^R_{min} &\leq v_0 - v_0^I \leq v^R_{max}\\
v_{min} \leq &\mu_{v_i}(\pmb{\hat{\lambda}},v_0) \leq v_{max} \ \ \forall i\in\{1,...,N\}
\end{split}
\end{equation}
where, $v_0$ is integrated into the optimization problem as a decision variable. Constraints are added to ensure that the control action and the expected nodal voltage magnitudes remain within permissible boundaries $[v_{min},v_{max}]$. Here, $v_0^I$ represents the initial setpoint value for $v_0$, and $[v^R_{min},v^R_{max}]$ is the permissible range of rate of change of voltage at the reference bus with respect to the initial voltage setpoint. To integrate $v_0$ into the problem, the expected nodal voltage magnitude squared values are written as a function of network parameters, expected available nodal active/reactive powers, and the optimization decision variables using \eqref{eq:mu}:
\begin{align}
&\left[
\begin{array}{c}
\mu_{v_1}\\
\vdots\\
\mu_{v_N}
\end{array}
\right]\approx\nonumber\\
&\left[
\begin{array}{ccc}
R_{11}P_1+X_{11}Q_1& \hdots &R_{1N}P_N + X_{1N}Q_N\\
\vdots & \ddots &\vdots\\
R_{N1}P_1+X_{N1}Q_1& \hdots &R_{NN}P_N+X_{NN}Q_N
\end{array}
\right]
\pmb{\hat{\lambda}} + \pmb{v_0}
\end{align}
where, $\pmb{v_0} = [v_0,...,v_0]^\top$.

Despite its convenient differentiable matrix-form formulation, the probabilistic tripping model introduces quadratic non-convex constraints into optimization problems. This challenge can be addressed using various relaxation techniques from the literature, such as semidefinite program (SDP) relaxation \cite{Ma2010}, second-order cone program (SOCP) relaxation \cite{Kim2001}, and parabolic relaxation \cite{Kheirandishfard2018}. To handle the non-convexity, these methods generally define an auxiliary matrix, $\Lambda = \pmb{\hat{\lambda}}\pmb{\hat{\lambda}}^\top$, which enables obtaining a convex surrogate for the original problem. For example, by applying parabolic relaxation, the constraints defined by the model are replaced with the following alternative constraints:
\begin{equation}
\label{eq:relax1}
\pmb{a_0} + (B-I_N)\pmb{\hat{\lambda}} + \left[
\begin{array}{c}
C_1\bullet \Lambda \\
\vdots\\
C_N\bullet \Lambda
\end{array}
\right] - \pmb{\epsilon}^+ \leq \pmb{0}
\end{equation}
\begin{equation}
\label{eq:relax2}
-\pmb{a_0} - (B-I_N)\pmb{\hat{\lambda}} - \left[
\begin{array}{c}
C_1\bullet \Lambda \\
\vdots\\
C_N\bullet \Lambda
\end{array}
\right] + \pmb{\epsilon}^- \leq \pmb{0}
\end{equation}
\begin{equation}
\label{eq:relax3}
\forall i,j:
\begin{cases}
\Lambda(i,i) + \Lambda(j,j) - 2\Lambda(i,j) \geq (\hat{\lambda}(i) - \hat{\lambda}(j))^2\\
\Lambda(i,i) + \Lambda(j,j) + 2\Lambda(i,j) \geq (\hat{\lambda}(i) + \hat{\lambda}(j))^2
\end{cases}
\end{equation}
where, $C_i\bullet \Lambda = \sum_{n=1}^{N}\sum_{m=1}^{N}\{C_i(n,m)\Lambda(n,m)\}$, $I_N$ is an $N\times N$ identity matrix, and $\pmb{\epsilon}^+$/$\pmb{\epsilon}^+$ are positive/negative small-valued slack variables that are used for transforming equality constraints defined by the model into two equivalent inequality constraints. The obtained inequalities \eqref{eq:relax1}-\eqref{eq:relax3} are convex constraints with respect to variables $\pmb{\hat{\lambda}}$ and $\Lambda$. 

\section{Numerical Experiments and Validation}\label{sec:num}
\begin{figure}
\centering
\includegraphics[width=0.9\columnwidth]{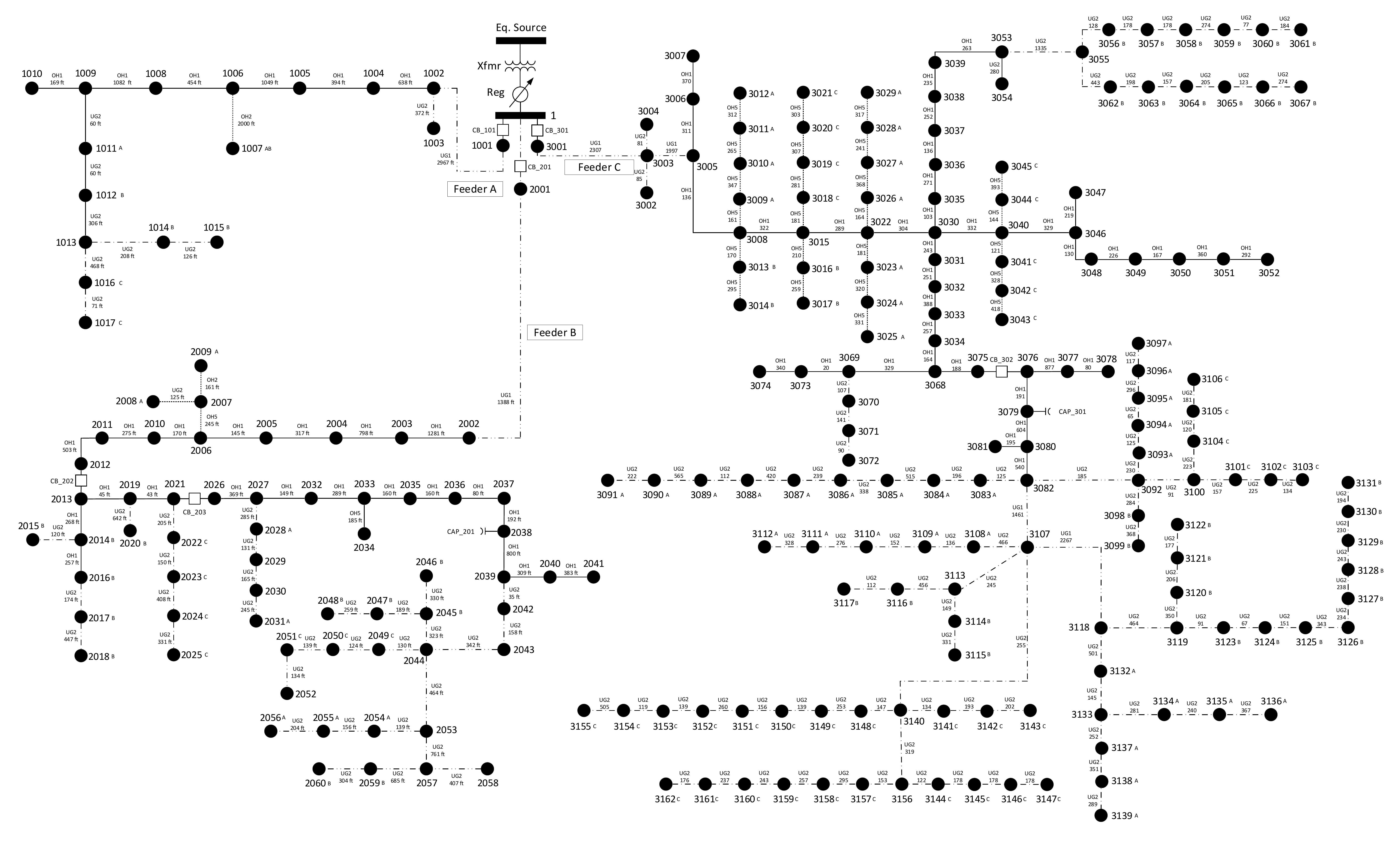}
\caption{Structure of the 240-node test system.}
\label{fig:network}
\end{figure}
\begin{figure}
\centering
\includegraphics[width=0.8\columnwidth]{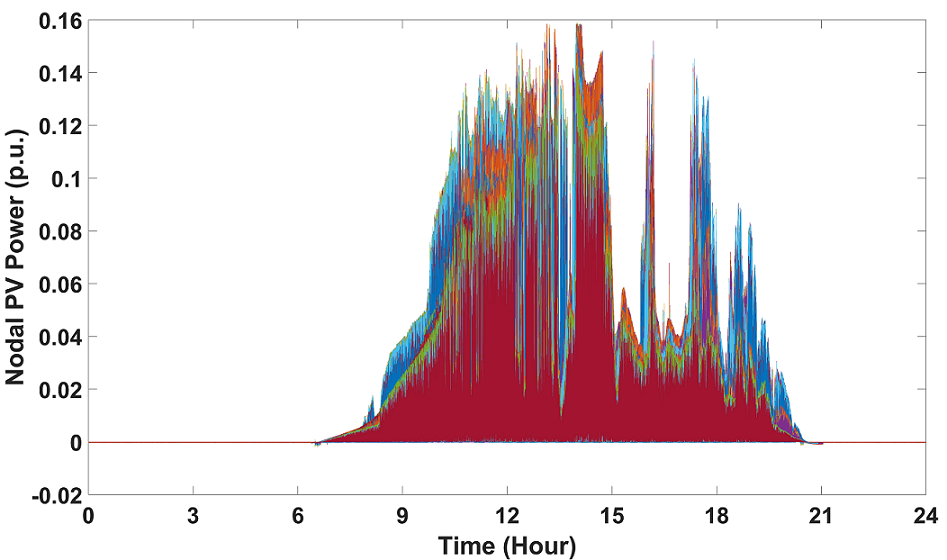}
\caption{Nodal PV outputs in the test system.}
\label{fig:solar}
\end{figure}
\begin{figure}
\centering
\includegraphics[width=0.8\columnwidth]{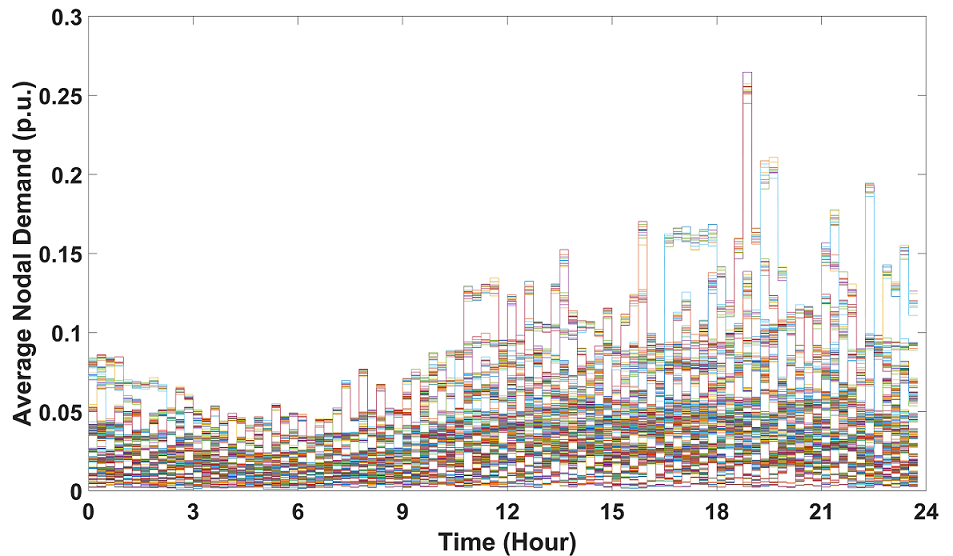}
\caption{Average 15-minute nodal consumption in the test system.}
\label{fig:load}
\end{figure}

Numerical experiments have been performed to validate the proposed probabilistic tripping model. In this, we have used real feeder model of an Iowa distribution system from our utility partner as shown in Fig. \ref{fig:network}. The network model in OpenDSS and detailed parameters are available online \cite{feeder}. To perform simulations we have used real solar and load data with 1-second time resolution from \cite{data}. Fig. \ref{fig:solar} shows the PV outputs at different nodes in the system for one day. Fig. \ref{fig:load} demonstrates 15-minute average nodal demand. The load/PV data have been randomly distributed across the three phases of the unbalanced grid at each node.

\begin{figure}
\centering
\subfloat[$\pmb{\lambda}$\label{fig:lam}]{
\includegraphics[width=0.4\textwidth]{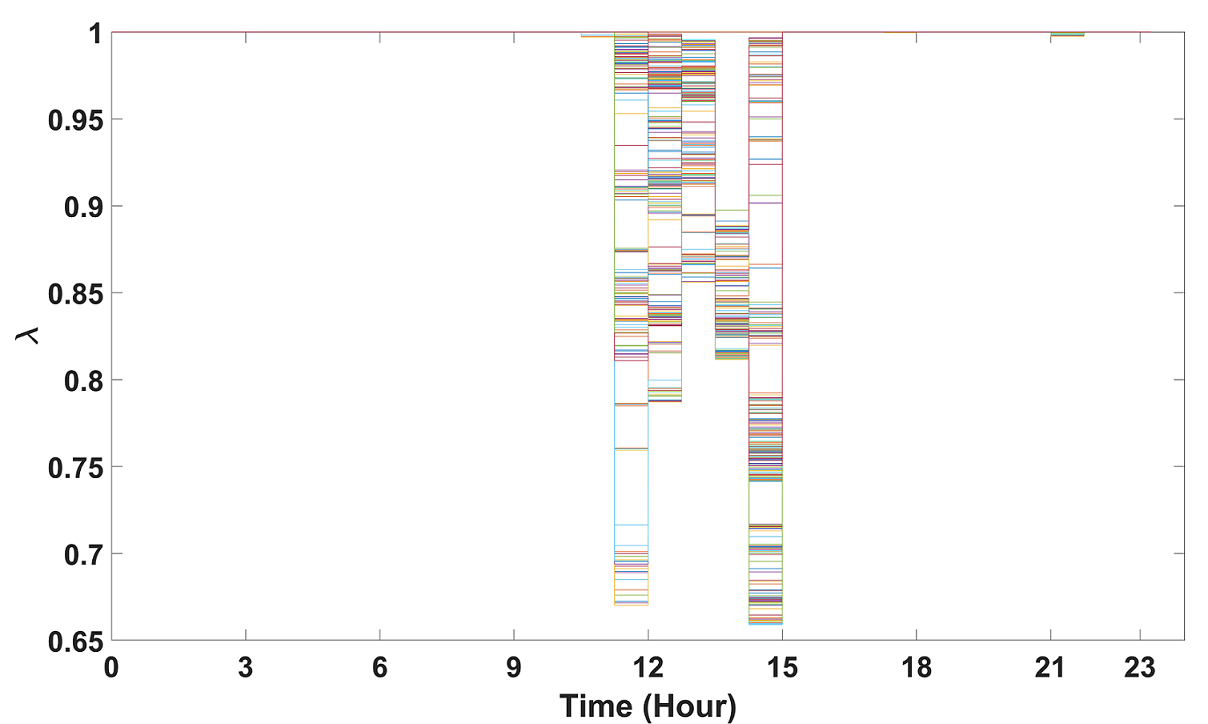}
}
\hfill
\subfloat[$S_p$ vs $\hat{S}_p$\label{fig:lower}]{
\includegraphics[width=0.4\textwidth]{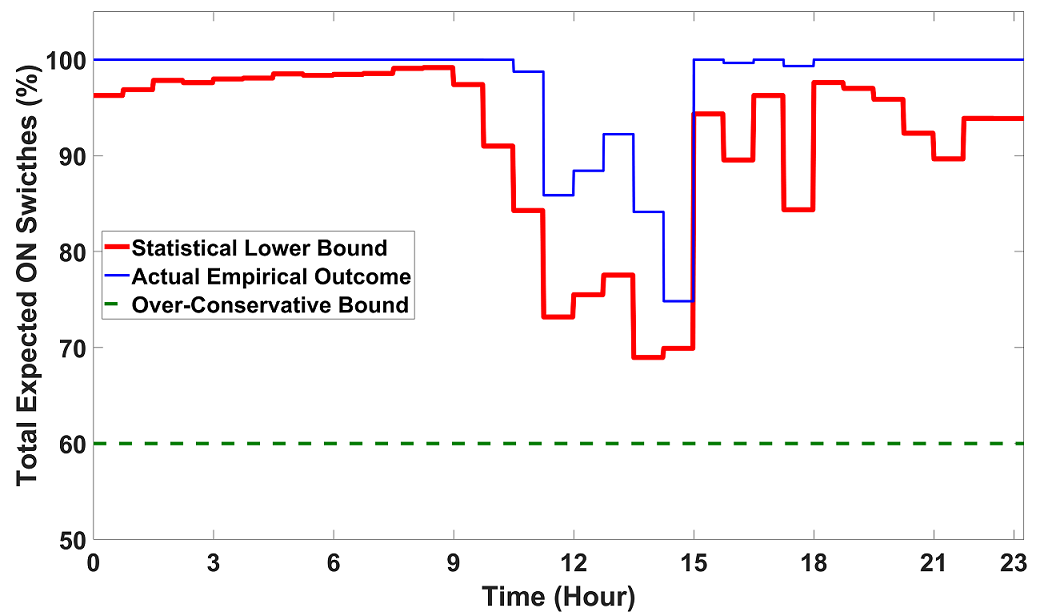}
}
\hfill
\subfloat[Aggregate expected PV power\label{fig:der}]{
\includegraphics[width=0.4\textwidth]{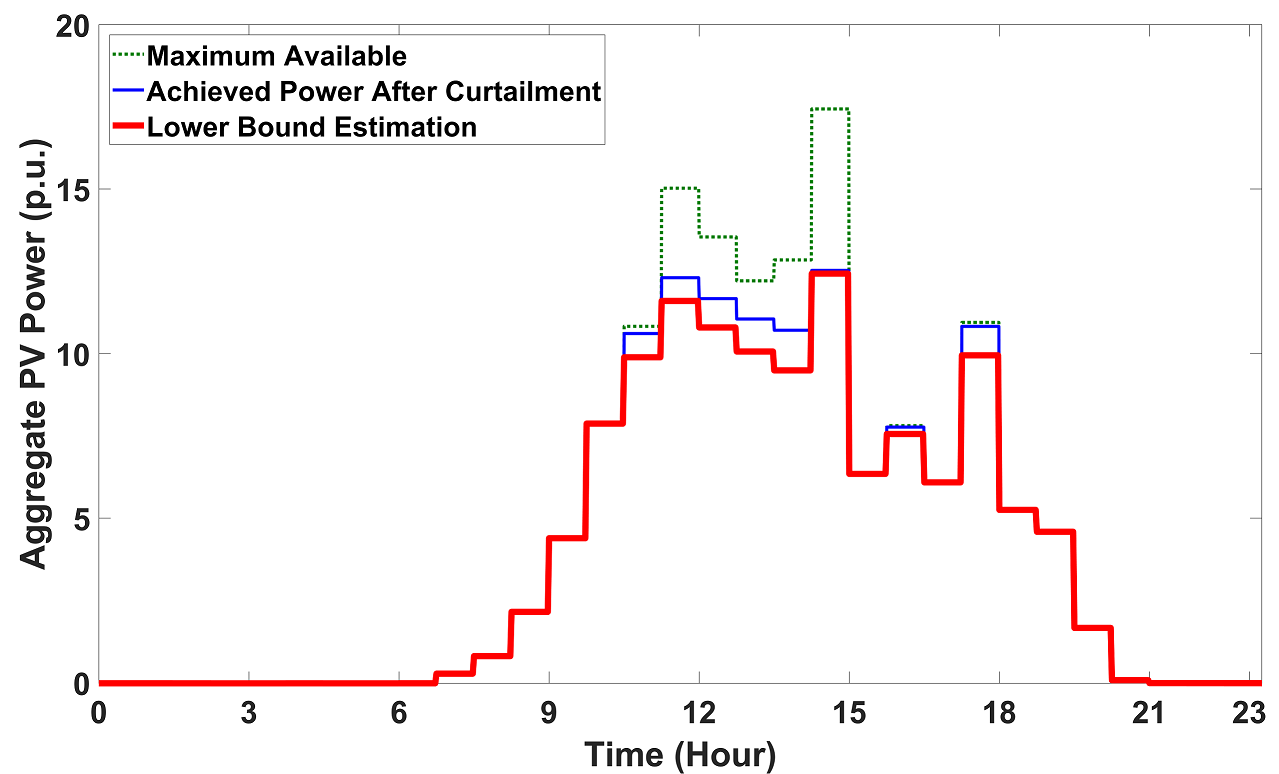}
}
\caption{Comparing the empirical and statistical lower bound solutions.}
\label{fig:lamS}
\end{figure}

\begin{figure}
\centering
\includegraphics[width=0.8\columnwidth]{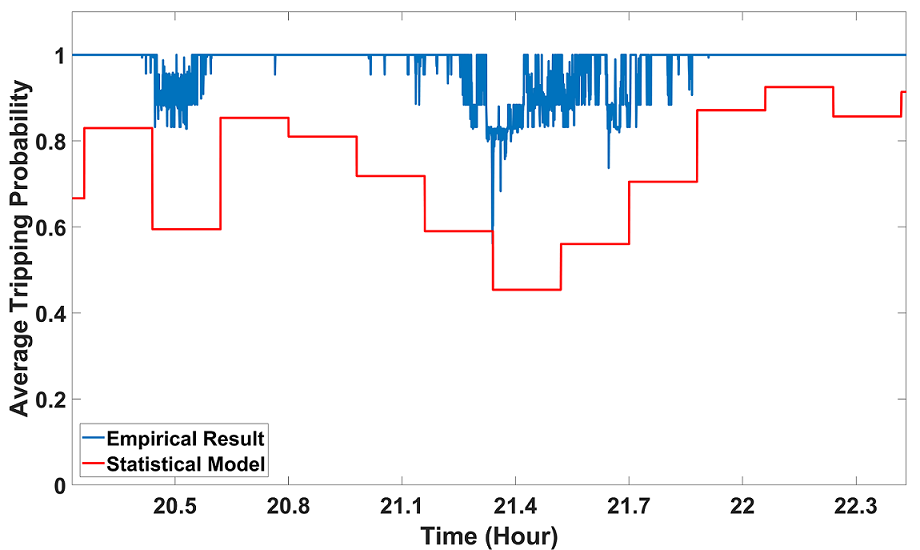}
\caption{Model performance for a case of heavy-loaded system and 15-minute empirical tripping probability assessment time window.}
\label{fig:under}
\end{figure}

To verify the performance of the proposed approximate statistical model, extensive time-series simulations were performed on the test system under various loading and solar generation scenarios over a course of day. Then, the real values of original micro-states, $\lambda_i$, were determined empirically over time windows of length $T = 60$ minutes. Intuitively, $\lambda_i$ serves as the \textit{ground truth} and roughly represents the portion of time that $s_i$ is ON during each time window:
\begin{equation}
\label{eq:lamempr}
\lambda_i(T) \approx \frac{\sum_{t = 1}^{T}s_i(t)}{T}
\end{equation}
Thus, we have two distinct time windows throughout numerical studies: a 1-second time step is used to perform high-resolution simulations, and a 1-hour time window is employed to obtain tripping statistics and empirically verify the performance of the proposed probabilistic model. Fig. \ref{fig:lam} demonstrates the empirical micro-states, $\lambda_i$, at different time intervals, which are determined by applying \eqref{eq:lamempr} to simulation outcomes. Based on the values of these micro-states, the empirical macro-state value is calculated at all time intervals, which represents the expected percentage of PV switches in ON state, i.e., $S_p(T) \approx \frac{\sum_{i = 1}^{N}\lambda_i(T)}{N}\times100$. Fig. \ref{fig:lower} compares the empirical macro-state value and the lower bound value constructed using solutions of \eqref{eq:Opt}. As can be seen, the solution from the probabilistic model actually represents a lower bound to the empirical macro-state obtained from simulations at all time windows, which corroborates the performance of the method. This figure also shows another lower bound obtained by simply using maximum PV capacities and assuming zero nodal consumption. However, as can be seen, this lower bound gives fixed over-conservative outcomes that do not reflect the true conditions of the system and have no correlation with the time-series PV/load data. Fig. \ref{fig:der} depicts the aggregate maximum available solar power (all switches ON at all time), empirical aggregate realized solar power from numerical simulations \eqref{eq:lamempr}, and solar power corresponding to solution of \eqref{eq:Opt}. As observed, the lower bound solution still holds and provides a conservative yet close estimation for the empirical achievable solar power outcome. Fig. \ref{fig:under} compares the empirical and model-based probabilities of inverter tripping in a heavy-loaded time interval. Unlike the previous case, these tripping probabilities are due to under-voltages. As can be observed, the model still provides a conservative lower bound on the probability of tripping. Note that the reason for higher levels of volatility in this figure is the shorter time window (15 minutes) used for assessing the empirical probability of tripping. 


\begin{figure}
\centering
\includegraphics[width=0.8\columnwidth]{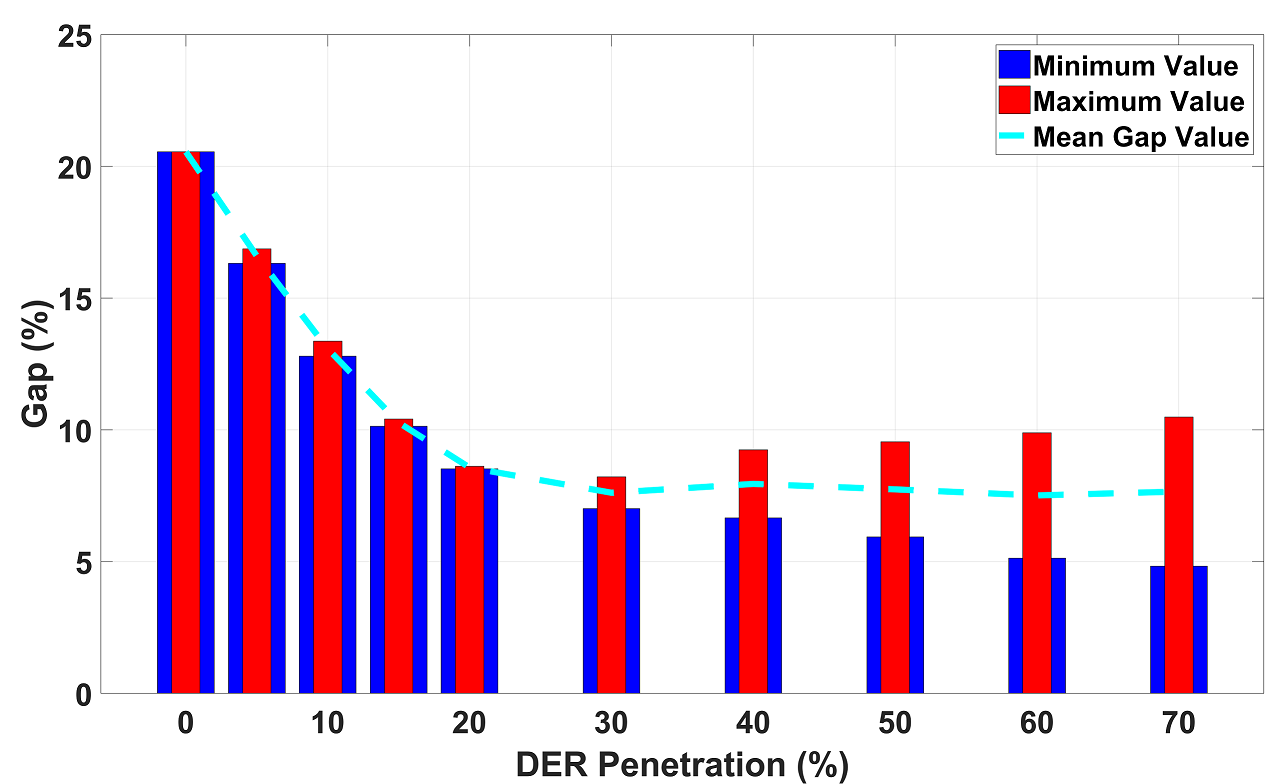}
\caption{Lower bound gap as a function of PV integration.}
\label{fig:dergap}
\end{figure}

The gap between the empirical macro-state obtained from numerical experiments and the proposed lower bound is an implicit function of PV penetration. Sensitivity analysis was performed to quantify the relationship between this gap and PV penetration percentage, as shown in Fig. \ref{fig:dergap}. Here, PV penetration is defined as the mean value of peak nodal solar power over peak nodal demand. The maximum, minimum, and mean values of the gap between the provided lower bound and the empirical macro-state is measured at various levels of PV penetration. As is observed in the figure, the optimistic value of the gap drops and eventually reaches $5\%$ as PV penetration increases, which indicates that the lower bound approaches the true macro-state value in grids with higher PV penetration. On the other hand, the maximum value of the gap shows an increase after a certain PV penetration level which points out to higher variations in solutions obtained from the probabilistic model. 

\begin{figure}
\centering
\includegraphics[width=0.8\columnwidth]{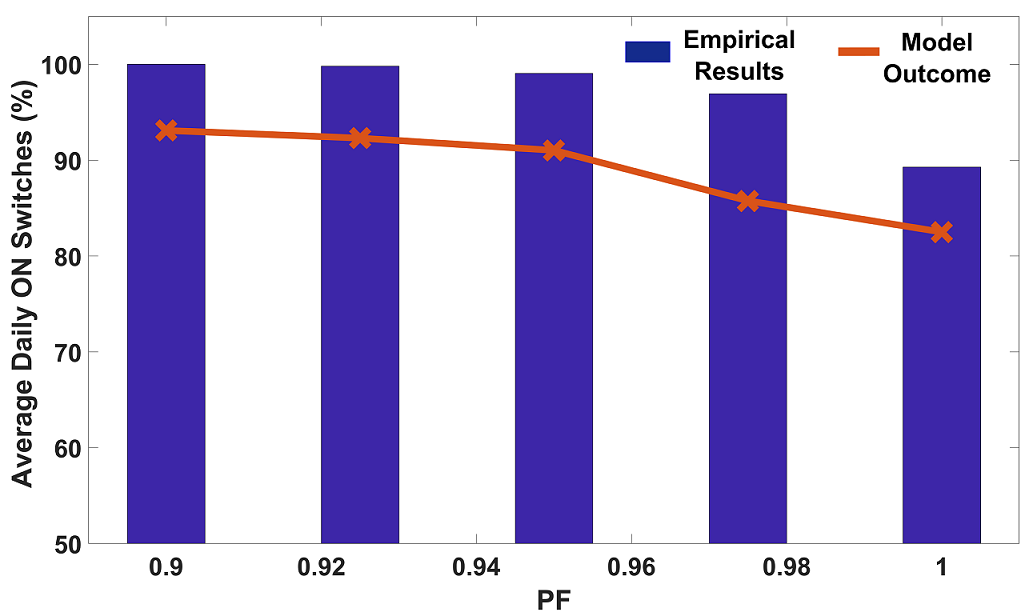}
\caption{Solar curtailment sensitivity to inverter control setpoints.}
\label{fig:pf}
\end{figure}

Fig. \ref{fig:pf} shows the overall daily solar curtailment levels, both empirical and the lower bound, as a function of changes in inverter control parameter. The inverters in the system are assumed to be controlled in constant power factor (PF) mode. As the reference PF setpoint increases and the system moves towards unity PF the voltage fluctuations increase, which leads to higher solar curtailment. This confirms previous observations in the literature \cite{Cheng2016}. Furthermore, our proposed probabilistic lower bound always slightly over-estimates the curtailment level, as expected correctly from the conservative estimator. 

\begin{figure}
\centering
\subfloat[Original and regulated $v_0$\label{fig:v0a}]{
\includegraphics[width=0.4\textwidth]{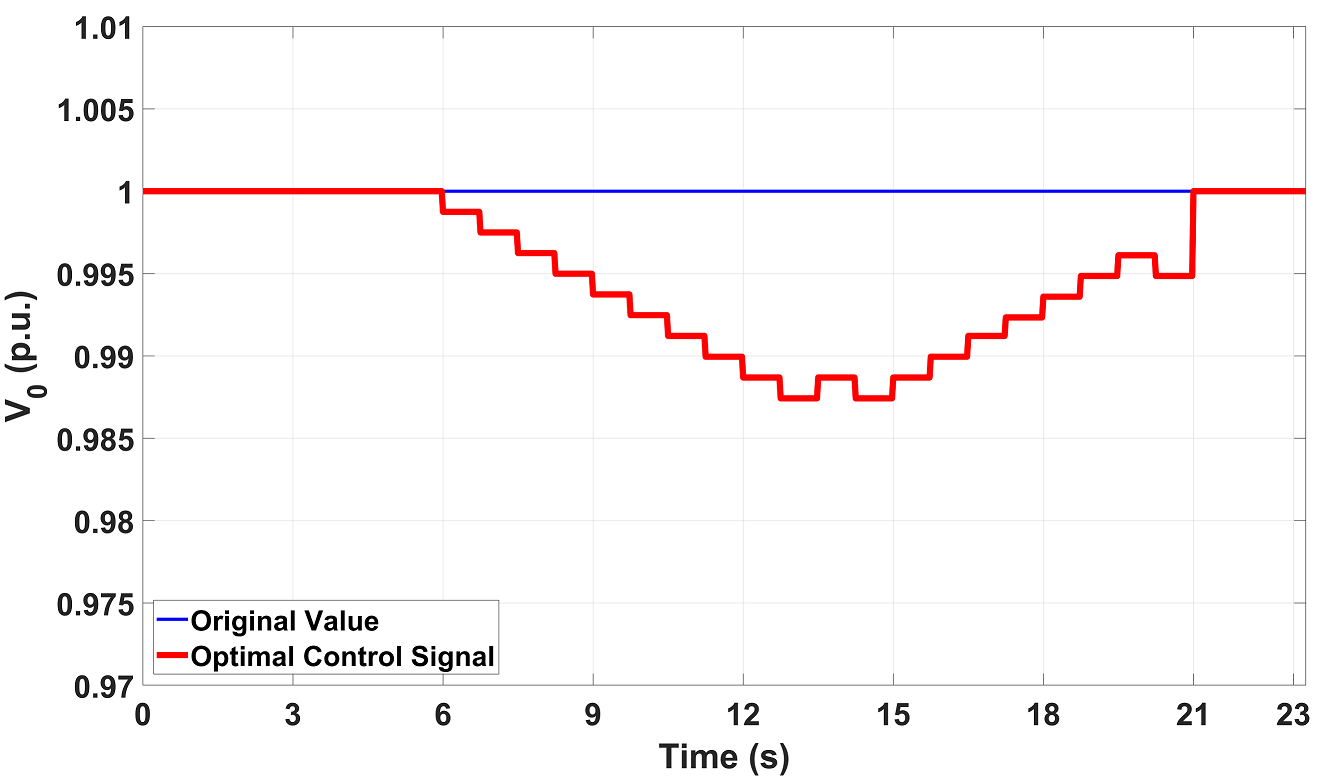}
}
\hfill
\subfloat[Macro-state under original and regulated $v_0$\label{fig:v0b}]{
\includegraphics[width=0.4\textwidth]{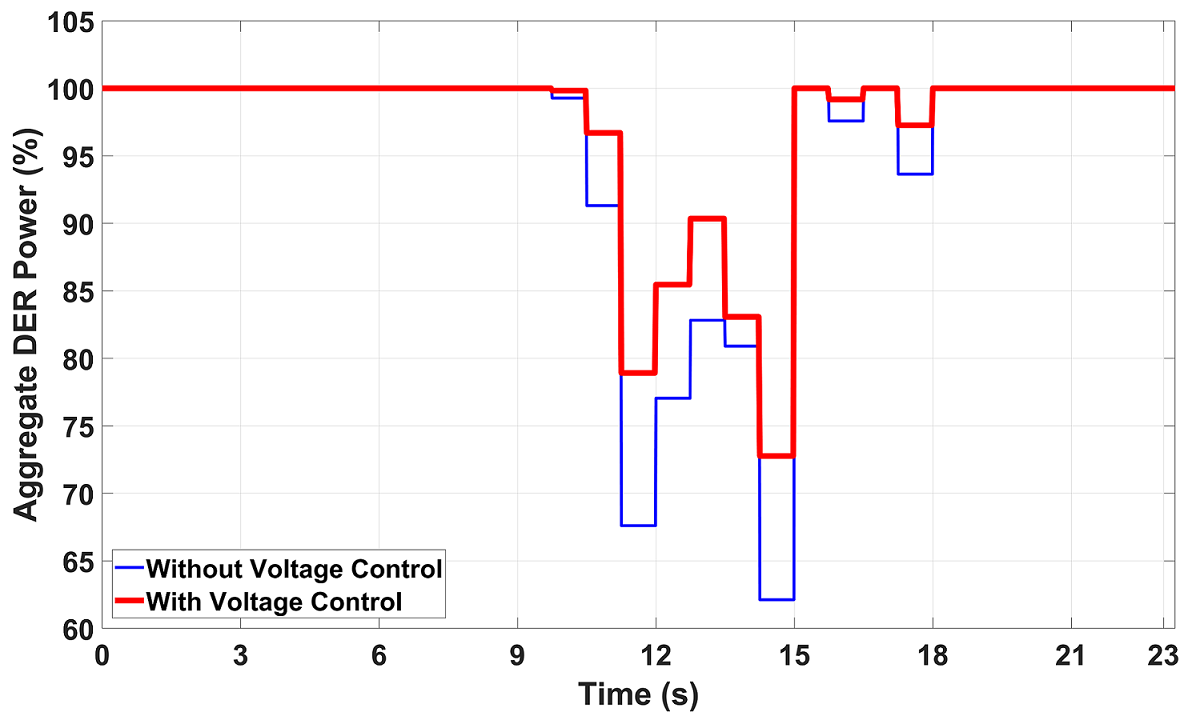}
}
\caption{Solar curtailment countermeasure design verification}
\label{fig:v0}
\end{figure}

Further tests were performed to corroborate the performance of countermeasure design strategy introduced in \eqref{eq:OptControl}. Fig. \ref{fig:v0a} shows the outcome of the optimization problem \eqref{eq:OptControl}, compared to a base case without any voltage regulation. As observed, $v_0$ is optimally decreased during solar-rich intervals to compensate for the increased voltage fluctuation levels. Fig. \ref{fig:v0b} compares the aggregate solar power injection values under the newly acquired $v_0$ values and the base case without voltage control. As can be seen, the obtained countermeasure has assisted significantly in mitigating the overall solar power curtailments during critical time intervals. 

\begin{figure}[t!]
\centering
\captionsetup{justification=centering}
\subfloat[Percentage of functional inverters\label{fig:UnderS}]{
\includegraphics[width=0.4\textwidth]{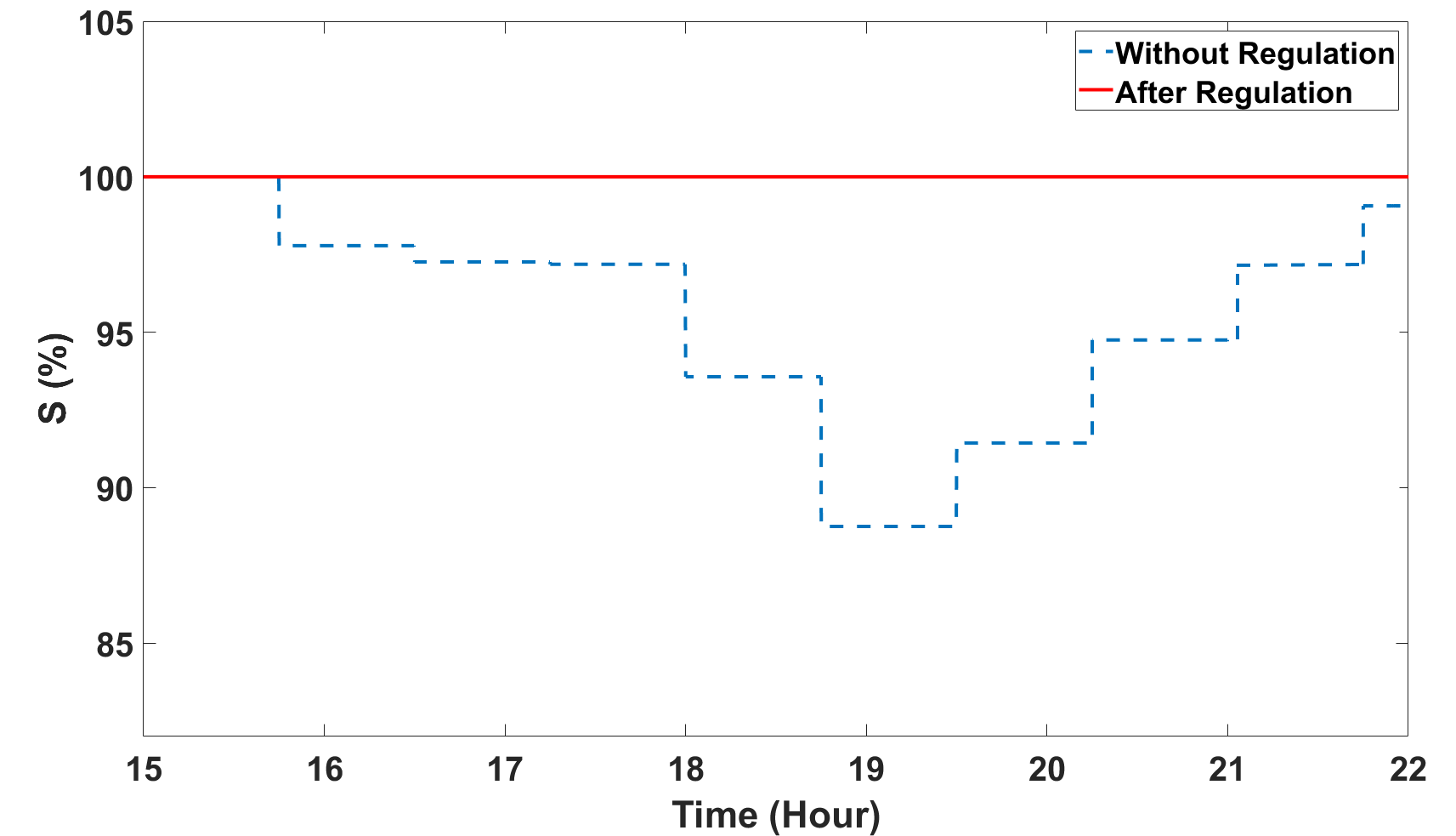}
}
\hfill
\subfloat[Aggregate nodal solar power\label{fig:UnderPV}]{
\includegraphics[width=0.4\textwidth]{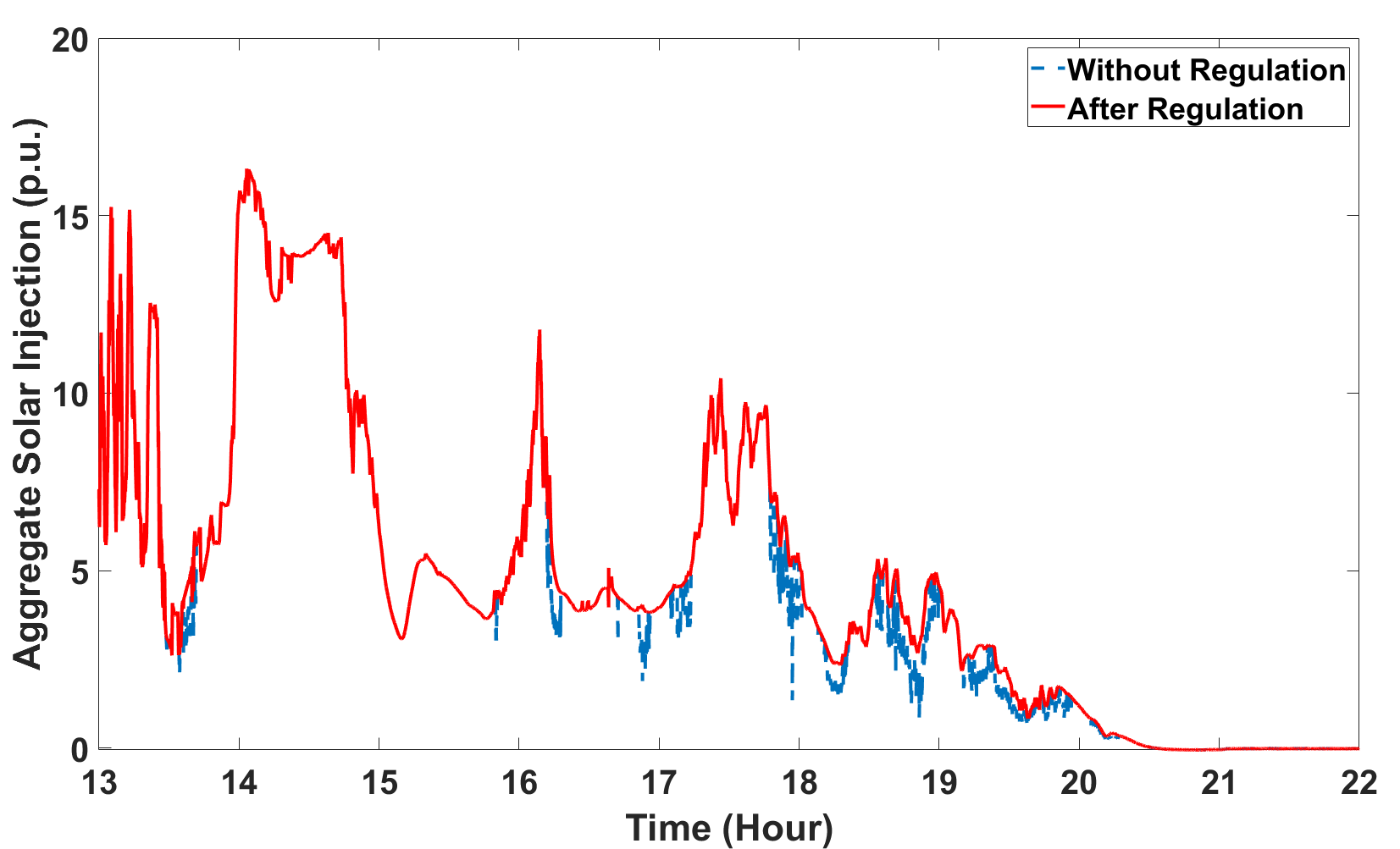}
}
\hfill
\subfloat[Average nodal voltage magnitude across the grid\label{fig:UnderV}]{
\includegraphics[width=0.4\textwidth]{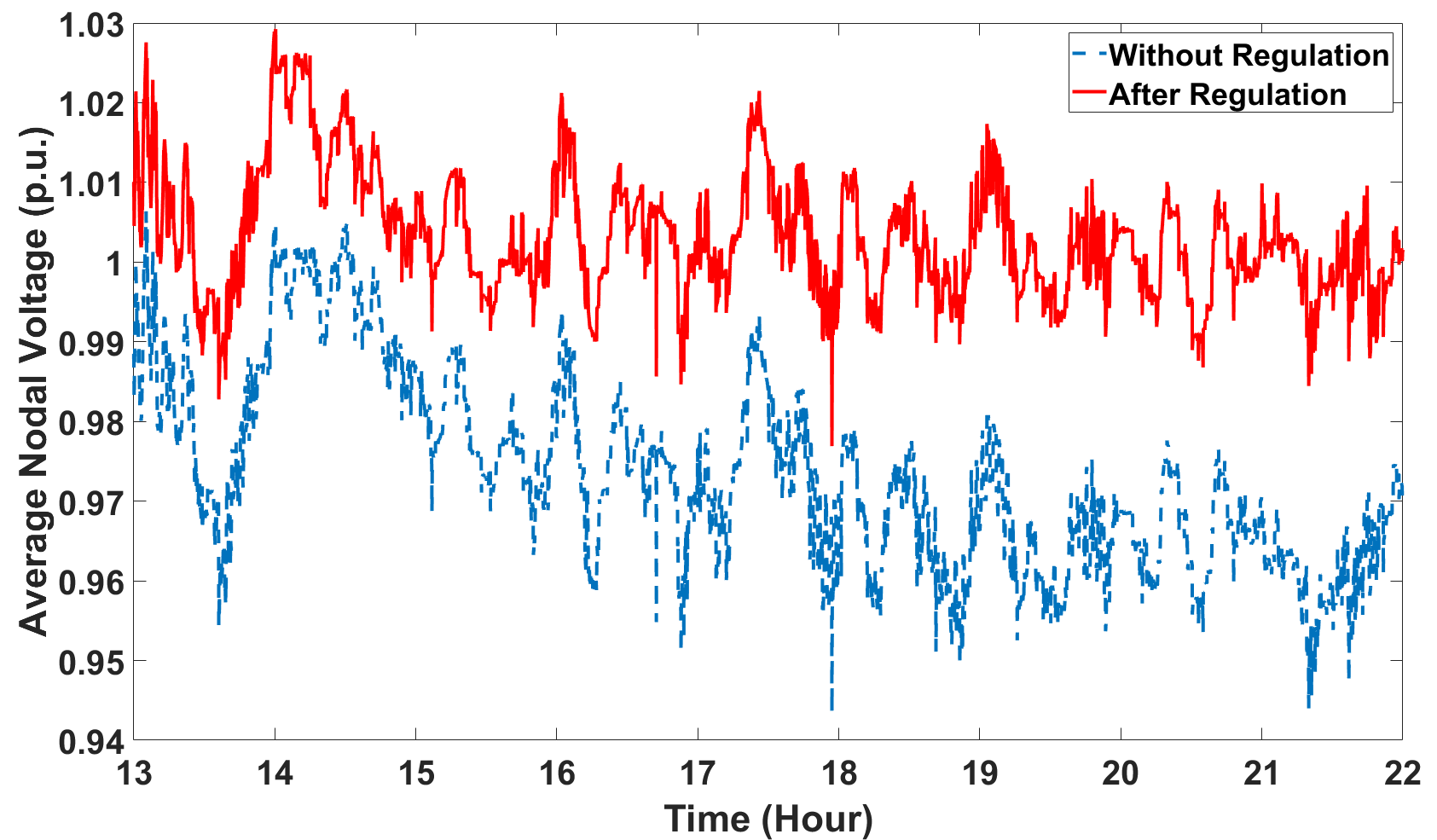}
}
\caption{An under-voltage case study.}
\label{fig:CaseUnder}
\end{figure}
We have performed another numerical experiment to analyse and verify the behavior of our tripping model during an under-voltage case study in a temporary heavy loading scenario in a weak grid under two strategies (see Fig. \ref{fig:CaseUnder}): (1) No voltage regulation is applied (baseline), and (2) Voltage regulation is applied with the objective of minimizing the average squared voltage deviations across the whole system, subject to linearized power flow equations and the proposed statistical tripping model. As can be seen in Fig. \ref{fig:UnderS}, under the baseline strategy (no voltage regulation) a portion of inverters (around 13\%) have tripped due to under-voltage protection during later hours of the day. This has resulted in a loss of renewable power injection in the grid (Fig. \ref{fig:UnderPV}). However, by applying voltage regulation using the proposed tripping model we have been able to maintain the voltages much closer to their nominal values (see Fig. \ref{fig:UnderV}) and prevent tripping events and loss of solar generation resources altogether. Note that Fig. \ref{fig:UnderV} shows the average value of nodal voltages across the whole system; thus, while most of the nodes maintain healthy voltage levels (as they should), the excessive loading on weak system lines under the baseline has resulted to a temporary voltage drop below inverters' protection activation threshold, which has engaged their under-voltage protection devices. This issue was mitigated using the deployed voltage regulation strategy that leverages our proposed statistical tripping model.

\begin{figure}
\centering
\includegraphics[width=0.9\columnwidth]{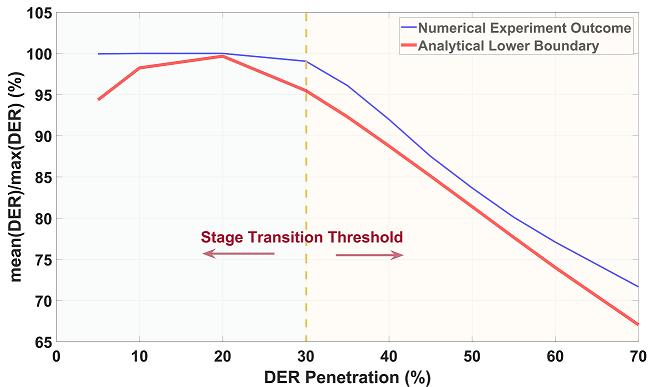}
\caption{Regime shift (stage transition) analysis.}
\label{fig:pen}
\end{figure}

Fig. \ref{fig:pen} demonstrates the average realized daily PV power ratio as a function of average PV penetration. As can be seen, the increasing penetration of solar has led to a regime shift after a certain threshold, from an initial state, in which the system shows almost no extensive tripping, to a new state, in which the average probability of solar curtailment steadily increases and extended tripping events can be expected. The existence of this threshold attests to a stage transition in the extent of switching events, which has been observed in other nonlinear systems as well \cite{Thurner2018}. Above the PV integration threshold, which is around 30\% for the test system, massive solar curtailment can be expected due to voltage fluctuations. It can be observed that the proposed statistical lower bound accurately tracks the behavior of the real system, and can be used to convey information on the whereabouts of the transition. The exact value of the regime shift threshold depends on many factors, including network topology and spatial-temporal distribution of loads/generators. 
\section{Conclusions}\label{sec:con}
In this paper, a probabilistic model of interdependent solar inverter tripping is presented to assess the risk of solar power curtailments due to voltage fluctuations in distribution grids. This model is developed using only the statistical properties of available load/PV active/reactive power. Numerical results on a real distribution feeder using real data successfully validate the estimated conservative lower bounds on inverter micro-states. Furthermore, it is demonstrated that the proposed model can be used for identifying regime shifts in tripping events and designing countermeasures to minimize risk of solar power curtailment. As a future research direction, we will explore integrating the more dynamic functions of inverter control and protection, including ride-through capabilities, \cite{NERC, AEMO2018, AEMO2017,Dozein2018} into the probabilistic tripping model. For example, the proposed statistical lower bound, which is based on Chebyshev's inequality, might become too conservative over short time windows if inverters' disturbance ride-through capabilities are activated. A less conservative lower bound that incorporates all aspects of inverter behavior will enable operators to monitor the sequence and transitions of tripping events, and mitigate potential cascading failure of resources.
\ifCLASSOPTIONcaptionsoff
  \newpage
\fi



\bibliographystyle{IEEEtran}
\bibliography{IEEEabrv,./bibtex/bib/main}
\end{document}